\documentclass[preprint2]{aastex}

\shorttitle{Temporal variations in X-ray solar flare loops}
\shortauthors{Jeffrey and Kontar}

\begin{document}

\title{Temporal variations of X-ray solar flare loops: length, corpulence, position, temperature, plasma pressure and spectra}

\author{Natasha L. S. Jeffrey and Eduard P. Kontar}
\affil{School of Physics \& Astronomy, University of Glasgow, G12 8QQ, Glasgow, Scotland, United Kingdom}

\email{n.jeffrey@physics.gla.ac.uk}

\begin{abstract}
The spatial and spectral properties of three solar flare coronal X-ray loops are studied before, during and after the peak X-ray emission. Using observations from the Ramaty High Energy Solar Spectroscopic Imager (RHESSI), we deduce the temporal changes in emitting X-ray length, corpulence, volume, position, number density and thermal pressure. We observe a decrease in the loop length, width and volume before the X-ray peak, and an increasing number density and thermal pressure. After the X-ray peak, volume increases and loop corpulence grows due to an increasing width. The volume variations are more pronounced than the position variations, often known as magnetic line contraction. We believe this is the first dedicated study of the temporal evolution of X-ray loop lengths and widths. Collectively, the observations also show for the first time three temporal phases given by peaks in temperature, X-ray emission and thermal pressure, with minimum volume coinciding with the X-ray peak. Although the volume of the flaring plasma decreases before the peak in X-ray emission, the relationship between temperature and volume does not support simple compressive heating in a collapsing magnetic trap model. Within a low $\beta$ plasma, shrinking loop widths perpendicular to the guiding field can be explained by squeezing the magnetic field threading the region. Plasma heating leads to chromospheric evaporation and growing number density, producing increasing thermal pressure and decreasing loop lengths as electrons interact at shorter distances and we believe after the X-ray peak, the increasing loop corpulence.
\end{abstract}


\section{Introduction}

During a solar flare, it is believed that magnetic energy in the corona is converted to other forms of energy;
causing the bulk motion and heating of surrounding plasma and accelerating particles up to non-thermal
energies. If the energy is released in a low density coronal region, then it is possible that the majority
of accelerated particles will leave the corona without losing energy collisionally until they reach the much higher
densities of the underlying chromosphere. Here the particles will interact with the surrounding plasma
and produce hard X-ray (HXR) bremsstrahlung emission. However, in most flares we also observe coronal X-ray
emission at or near to the top of a loop \citep[see][as recent reviews]{2011SSRv..159..107H, 2011SSRv..159..301K},
although with an interesting exception \citep{2011ApJ...731L..19F}.
Since bremsstrahlung emission grows with density, the low coronal densities can often make the observation
of coronal loop emission challenging, especially if it is masked by brighter chromospheric
X-ray emission \citep{2008ApJ...673.1181K}. However, the observation of HXR coronal sources with instruments such
as the Ramaty High Energy Solar Spectrometer Imager (RHESSI) \citep{2002SoPh..210....3L} are important since these X-ray loops
are much closer to or may even contain the actual sites of energy release and primary electron
acceleration \citep{2008ApJ...673..576X,2011ApJ...730L..22K,2012ApJ...754..103B,2012ApJ...755...32G}.
RHESSI creates X-ray images using Rotating Modulation Collimators (RMCs) \citep{2002SoPh..210...61H,2007SoPh..240..241S}.
As the spacecraft spins, the grids placed in front of the detectors produce a modulation curve of the X-ray intensity. From this we can
construct X-ray visibilities, which are equivalent to the visibilities used in radio astronomy \citep{2007SoPh..240..241S}. In the last few years,
an imaging technique known as visibility forward fitting (VIS FWDFIT) has been successfully used to study the sizes of chromospheric
footpoint sources \citep{2008A&A...489L..57K,2010ApJ...717..250K,2011ApJ...735...42B} and inferred that the cross-sectional sizes
of HXR sources decrease with energy.
This technique has the advantage over other imaging algorithms such as Clean and Pixon etc. because it provides us with relatively
accurate position and size measurements from the moments of the X-ray distribution,
plus an error for each value. However, X-ray visibility fitting also has the disadvantage of limiting 
the number of sources and their shapes that can be studied using this technique.
Therefore, careful comparisons with other imaging algorithms are required when using VIS FWDFIT.

\begin{deluxetable}{c|ccccc}
\tabletypesize{\normalsize}
\tablecaption{Table showing the main parameters of each flare.}
\tablewidth{0pt}
\tablehead{\colhead{} & \colhead{   GOES Class} & \colhead{Date} & \colhead{Obs. time} & \colhead{Peak time (10 keV)} & \colhead{Footpoints (30-40 keV)} }
\startdata
Flare 1 &  M3.0 & 23-August-2005 & 14:22:00-14:40:00 & 14:30:00 & 14:36:00 onwards\\
\hline
Flare 2 &  M4.1 & 14/15-April-2002 &23:58:00-00:20:00 & 00:12:00 & 00:05:00 onwards\\
\hline
Flare 3 & M2.6 & 21-May-2004 & 23:42:00-23:58:00 & 23:50:00 & 23:42:00 onwards\\
\enddata
\label{tab:tab_paras}
\end{deluxetable}

Recently, a number of studies have used visibility forward fitting
to study the sizes of coronal X-ray loop sources. \cite{2008ApJ...673..576X} and \cite{2012ApJ...755...32G} investigated how the lengths of X-ray coronal loops,
that is, the direction parallel to the guiding magnetic field, varied with photon energy during the X-ray impulsive stage.
For a number of different flares, they found that the loop lengths increased with energy with a form consistent with
that of an initial extended acceleration region within the loop itself, plus an additional length proportional to the photon
energy squared. This is due to higher energy electrons moving further through the plasma before interacting.
\cite{2011ApJ...730L..22K} studied the length and width changes of one X-ray coronal loop with photon energy.
They found that both the length and width increased with photon energy, with the loop width increasing proportionally to the photon energy.
Increases in loop width, or corpulence, are more difficult to explain since the electrons are bound to the guiding field and cross field transport
should be negligible. \cite{2011ApJ...730L..22K} and \cite{2011A&A...535A..18B} inferred that the width increase could be due to the magnetic diffusion
of field lines perpendicular to the direction of the field, caused by the presence of magnetic turbulence within the loop.
Although, these observations indicate the usefulness of coronal loop spatial properties, the spatial evolution of loop sources has not yet
been fully explored. Loop positions have been studied before, during and after the impulsive phase of the flare.
\cite{1996ApJ...459..330F} observed in soft X-rays the changing locations of post flare loops,
which was interpreted as the decrease in height that open field lines undergo after they have
reconnected to form closed loops.
They studied two long duration events near the limb and found that field loop shrinkage did occur and matched the shrinkage predicted by a simple model of the reconnecting field but found overall the entire flare loop system grew with time.
\cite{2003ApJ...596L.251S}, \cite{2004ApJ...612..546S}, \cite{2006A&A...446..675V} and \cite{2009ApJ...706.1438J} all noted a decrease in the altitude of loop top sources
during the impulsive phase of the flare, until the peak X-ray emission and an increase in altitude after the impulsive phase.
\cite{2003ApJ...596L.251S} also found evidence for an above the loop top source and interpreted the situation as the formation
of a reconnection current sheet between the loop top source and the higher coronal source.
\cite{2006A&A...446..675V} interpreted the decrease as a collapsing magnetic trap \citep{1997ApJ...485..859S,2004A&A...419.1159K}.
The contraction and expansion of the loop source has also been observed in other wavelengths of EUV \citep{2009ApJ...696..121L,2009ApJ...706.1438J}
and radio \citep{2005ApJ...629L.137L,2010ApJ...724..171R}.  \cite{2010ApJ...724..171R} found both changes in the radio loop span and height with time.
More recently, \cite{2012ApJ...749...85G} looked for evidence of collapsing fields using Solar Dynamics Observatory (SDO/AIA and HMI) observations.
The loops rose slowly and then moved into a collapse phase during the impulsive phase of the flare, where the loop tops contracted. Lower loops contracted earlier than higher loops and the loop contraction was interpreted as a reduction of magnetic energy as the system relaxed to a state of lower energy, i.e. relaxation theory \citep{1974PhRvL..33.1139T}.

In this paper, we study three flares with coronal X-ray loops to find how the emission lengths, widths and positions change with time at three energy ranges
between 10-25 keV. From each flare spectra, we will also investigate how parameters such as emission measure and plasma temperature vary with time.
Using a combination of imaging and spectroscopy parameters, we can infer how the X-ray loop volume, number density, thermal pressure
and energy density vary during the time evolution of the flare, allowing us to build a fuller picture and propose some explanations describing the processes occurring
within the coronal loops. 

\section{Chosen events with coronal X-ray emission}

The three events studied are: 23rd August 2005 from 14:22:00 (Flare 1),
14th/15th April 2002 from 23:58:00 (Flare 2) and 21st May 2004 from 23:40:00 (Flare 3).
All three flares share similar characteristics: GOES M-class flares with similar lightcurves, strong coronal X-ray loop top emission
and only relatively weak footpoint emission. Since the aim of our study is to examine how the properties of coronal loop emission
vary with time, these events were chosen as they show loop emission throughout the rise, peak and decay stages
of X-ray emission and their spatial properties have been previously studied by \cite{2008ApJ...673..576X} and \cite{2011ApJ...730L..22K}.
The coronal X-ray emission during each flare appears up to $\sim 25$ keV and each source is a simple loop-like shape
connecting HXR footpoints at 30-40 keV during certain time intervals. The main parameters of each flare: GOES class, date,
observation time, peak time at 10 keV and the time of footpoint appearance are given
in Table~\ref{tab:tab_paras}. Length variations of each of these coronal loops with photon energy were studied by \cite{2008ApJ...673..576X}.
Length and width changes with photon energy for Flare 2 were also studied by \cite{2011ApJ...730L..22K}. It should be noted that Flare 1
and Flare 3 show similar results as Flare 2 in \cite{2011ApJ...730L..22K}, where both the length and corpulence increase with
energy as $\sim\epsilon^{2}$ and $\sim\epsilon$ respectively but this paper will concentrate on the size, position and spectral parameter
changes with time. The lightcurves for each event are shown in Figure \ref{fig:is_paras} (top row). Flares 1 and 3 have similar lightcurves;
a simple shape with one peak. The lightcurve for Flare 1 is shown in Figure \ref{fig:is_paras} (top row, left plot). Our study of this event begins
at 14:22:00. At this time, emission from the 10 to 20 keV energy bands are slowly rising and reach a peak at $\sim$14:30:00.
After this point, there is a gradual decrease in X-ray emission, which can be observed up until $\sim$14:50:00, where RHESSI enters into a night phase.
In the 20-40 keV band we see a series of peaks between our observation range of 14:22:00 and 14:40:00. The lightcurve for Flare 3 is shown
in Figure \ref{fig:is_paras} (top, right plot). From the start of our study at 23:40:00, the X-ray emission is rising from 6-50 keV
and peaks around 23:50:00. After this peak, the X-ray emission decreases. The lightcurve for Flare 2 is more complex and is shown
in Figure \ref{fig:is_paras} (top row, middle plot). During the observational time there are two main peaks
in the lightcurve at $\sim$00:03:00 and $\sim$00:12:00, possibly more peaks, which are easiest
to see in the 20-40 keV energy range.

\begin{figure*}
\centering
\includegraphics[scale=.55]{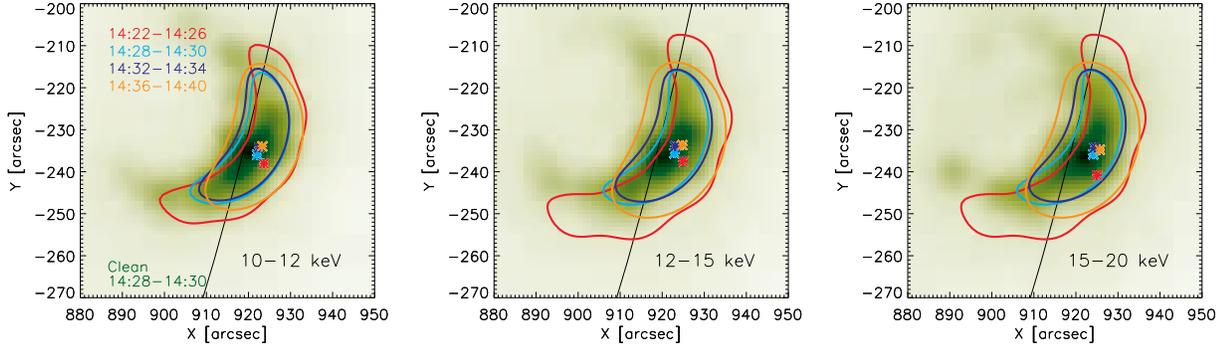}
\caption{23-August-2005 (Flare 1). Clean image (green) at 14:28:00-14:30:00 and over-plotted VIS FWDFIT contours (50\% maximum intensity) 
at four selected times for 10-12 keV (left), 12-15 keV (middle) and 15-20 keV (right). The asterisks denote the loop position for each selected time. 
The cyan VIS FWDFIT contour matches the time of the Clean image.}
\label{fig:img_flare1}
\end{figure*}
\begin{figure*}

\centering
\includegraphics[scale=.55]{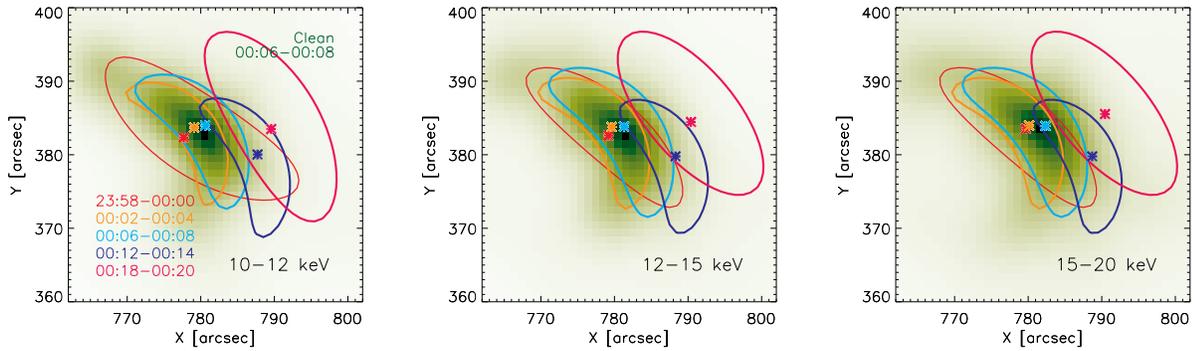}
\caption{As Figure \ref{fig:img_flare1} but for Flare 2, 14th April 2002.}
\label{fig:img_flare2}
\end{figure*}
\begin{figure*}

\centering
\includegraphics[scale=.55]{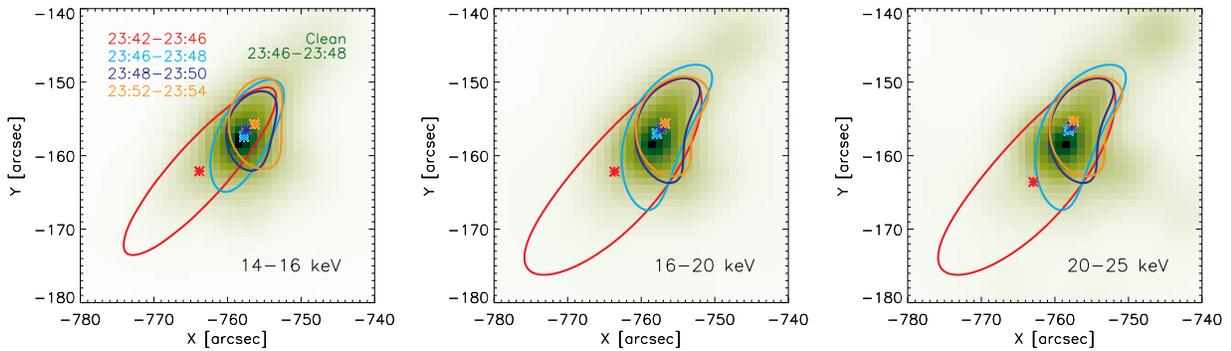}
\caption{As Figure \ref{fig:img_flare1} but for Flare 3,  21st May 2004. Flare 3 uses the energy ranges of 14-16 keV (left), 16-20 keV (middle) and 20-25 keV (right).}
\label{fig:img_flare3}
\end{figure*}

\section{Spatial and spectral changes with time}

\begin{figure*}
\vspace{-10 mm}
\includegraphics[scale=.35]{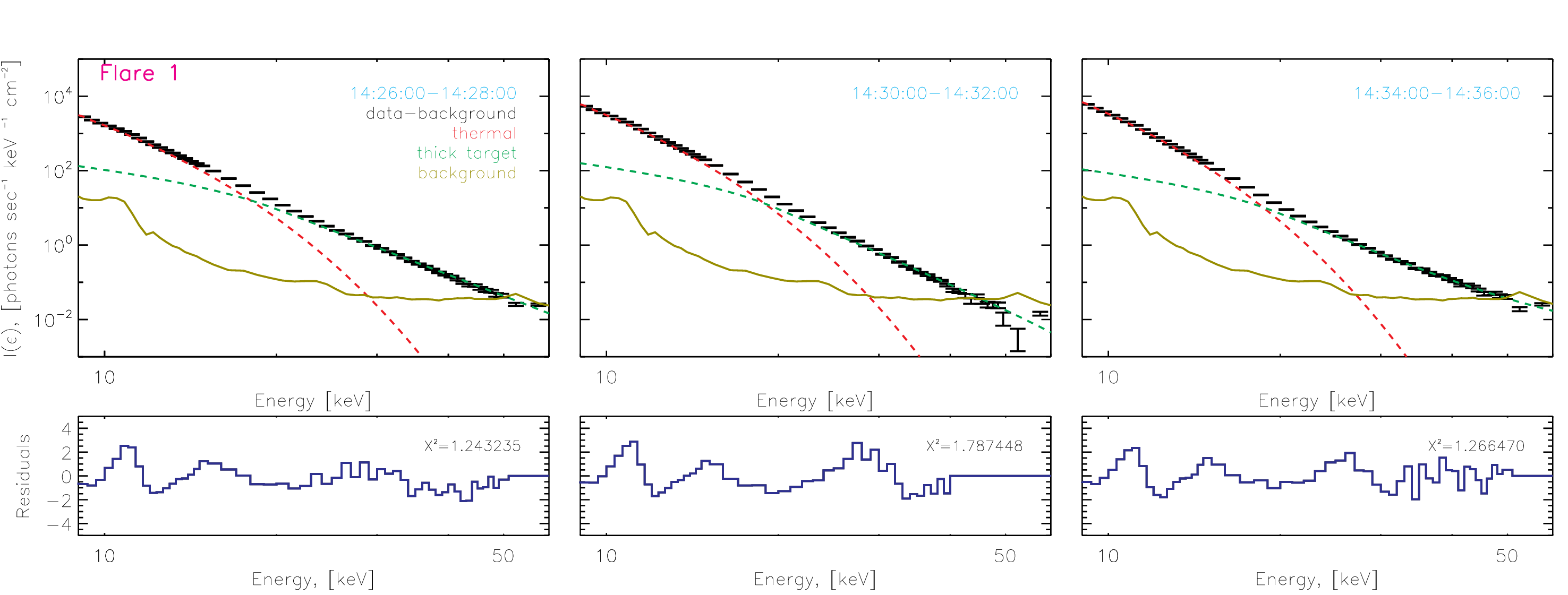}
\includegraphics[scale=.35]{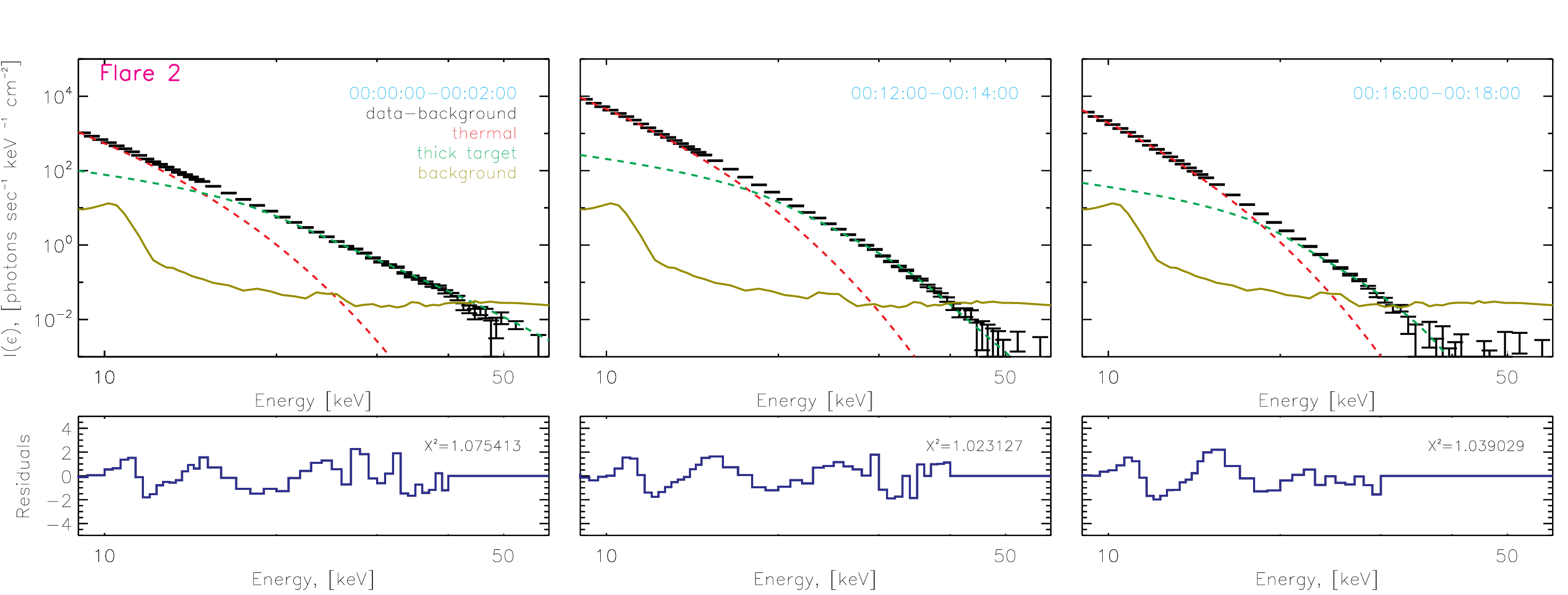}
\includegraphics[scale=.35]{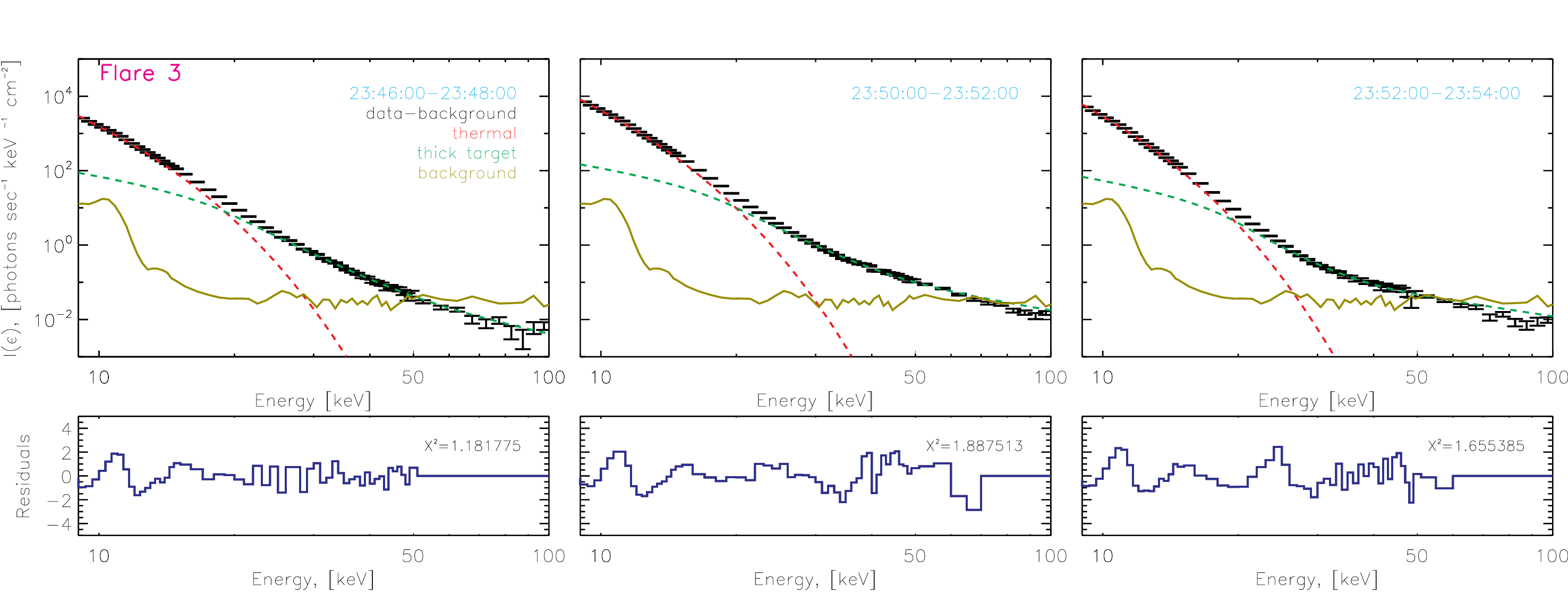}
\caption{Spectra for Flares 1 (top row), 2 (middle row) and 3 (bottom row), at three chosen imaging time bins (during the X-ray rise, peak and decay stages). Black = data - background, Olive = background, red = thermal fit, green = thick target bremsstrahlung fit. The residuals are plotted below each plot.}
\label{fig:spectra}
\end{figure*}

Each event was studied using the imaging algorithms of Clean \citep{1974A&AS...15..417H, 2002SoPh..210...61H},
Pixon \citep{1993PASP..105..630P,1996ApJ...466..585M}  and VIS FWDFIT \citep{2002SoPh..210...61H,2007SoPh..240..241S}.
Firstly, we studied each of our three events using only Clean and Pixon to confirm the loop shape of each coronal source
and find the energy ranges over which a coronal source was present in each flare. Once we were confident that our
chosen events only had a simple loop shape, we studied each event using VIS FWDFIT. This is important since the coronal source
must have a loop-like shape so that VIS FWDFIT can effectively fit a curved elliptical gaussian (loop) to the X-ray visibilities
of the event and give realistic estimates with errors for the source parameters. VIS FWDFIT provides us with loop length
FWHM (full width half maximum), loop width FWHM and the $(x,y)$ centroid position of the loop.
It should also be noted that VIS FWDFIT gives the mean position of the loop shape, not the loop top centroid position,
that is, the central position at the top of the loop. The coordinates for loop top position were obtained by extracting
the coordinates of the central circular gaussian that VIS FWDFIT uses to create the final loop shape along
with a set of other circular gaussians placed along the length of the loop. This is important as a loop that is very
curved and approaching the shape of a ring will pull the shape centroid towards the ends of the loop,
often masking small changes in loop position with time or energy. This was especially significant for the large loop shape
of the 23rd August 2005 event (Flare 1). In order to study changes in time at a specific energy range, we imaged the coronal
sources of Flares 1, 2 and 3 over two minute intervals where possible and four minute intervals where the count rates were lower.
The exact time bins used for each flare are shown in Figures \ref{fig:is_paras} and \ref{fig:com_paras}.
The energy ranges of 10-12 keV, 12-15 keV and 15-20 keV were chosen for Flare 1 and Flare 2  and the energy
ranges of 14-16 keV, 16-20 keV and 20-25 keV were chosen for Flare 3. These energy ranges were chosen since VIS FWDFIT with a loop
produced clear results with small errors at these energy ranges. Images for Flares 1, 2 and 3 are shown
in Figures \ref{fig:img_flare1}, \ref{fig:img_flare2} and \ref{fig:img_flare3} respectively. Figure \ref{fig:img_flare1} shows that Flare 1
is a limb event while Figures \ref{fig:img_flare2} and \ref{fig:img_flare3} show that both Flare 2 and Flare 3 are disk events
(the main parameters of each flare are given in Table~\ref{tab:tab_paras}). Figures \ref{fig:img_flare1}, \ref{fig:img_flare2}
and \ref{fig:img_flare3} plot a background Clean image of the coronal source for each flare at one chosen time interval
and over plot VIS FWDFIT contours for selected time intervals, one corresponding to the same time interval as the Clean image.
Comparing the shape and size of the VIS FWDFIT contour with the Clean background image at the selected time interval
for each flare shows good agreement between both algorithms. The Clean images here use a sigma beam width of 1.8.
It should be noted that we also found the standard deviation from image cuts through the images of Clean, Pixon and VIS FWDFIT
and it was found that a Clean sigma beam width of $\sim$3.0 showed the best agreement with the standard deviations of Pixon
and VIS FWDFIT. In terms of imaging, VIS FWDFIT also agrees well with the Clean and Pixon algorithms at all other times
not shown in this paper. From Figure \ref{fig:img_flare3}, it should be noted that the results for Flare 3
are probably less reliable than those of Flare 1 and Flare 2. From the position of the footpoints in the Clean image,
it appears as through the southern `loop leg' is tucked underneath the observer's line of sight.
This means that it is harder for VIS FWDFIT with a loop to fit it with a correctly shaped loop and usually fits it with a loop
that is slightly too large or departs from a loop-shape. For each of the imaging time intervals for Flares 1, 2 and 3, energy spectra
were created. The spectra of each flare at each time interval were fitted with a thermal component and a non-thermal
component corresponding to thick target bremsstrahlung. The thermal fits provide us with the emission measure, $EM$, and
the plasma temperature, $T$. We note all three flares show the emergence of footpoints in the 30-40 keV energy range.
Spectra for Flares 1, 2 and 3 are shown in Figure \ref{fig:spectra} at three selected time bins corresponding to a rise, peak
and decay stage of X-ray emission for each flare.

Figure \ref{fig:is_paras} shows the imaging parameters for Flare 1 (column 1), 2 (column 2) and 3 (column 3):
loop width FWHM (row 2), loop length FWHM (row 3) and loop-top radial position (row 4) for each imaging
energy band. Above these plots, the lightcurve is plotted for each of the imaging energy bands to allow
comparison with changes in the spatial parameters. The dashed lines drawn in each plot represent
the time interval over which the peak X-ray emission occurs for our three energy bands.
For each flare, in general, peaks in the lightcurve represent changes of width, length and source position
parameters with time. The last two rows of Figure \ref{fig:is_paras} plot how the spectroscopic parameters
vary with time for each flare. In general, for all three events, the emission measure rises throughout the observation time,
either slowing or decreasing slightly during the last few minutes for each event.
For each event, the plasma temperature peaks before the peak in X-ray emission
and then slowly decreases after this point, as noted by \cite{1978ApJ...220.1137A}. The plasma temperature decreases much
slower than if it were decreasing by conduction only, suggesting prolonged energy release at the later decay stages of the flare.

\subsection{Loop corpulence}
For all three flares, we see a general pattern for emitting loop width changes with time.
Before a peak in X-ray emission, the source width for each flare tends to decrease
and after a peak in X-ray emission, the source width for each flare increases.
It is also notable that the rates of both width expansion and contraction (width change in time) are
approximately the same as can be seen in Figure \ref{fig:dvwl}.

{\it Flare 1-} For all energies plotted, the source width decreases until the peak X-ray emission at $\sim$14:30:00-14:32:00 UT
and then increases after this point. The largest change occurs for the highest energy of 15-20 keV.
This falls from $14.5''$ to $11.5''$ at 14:28:00-14:30:00 and then increases until 14:40:00
where the width peaks at $\sim 20''$, producing a final larger loop corpulence than seen at the beginning of our observational time.

{\it Flare 2-} For the energies of 10-12 keV and 12-15 keV, the loop width decreases before the first peak in X-ray emission
at $\sim$00:03:00. After this point the width increases before dropping at 00:12:00-00:13:00 where there is another X-ray peak
in the lightcurve. After this peak, the width continues to grow again. Before 00:03:00, the width of the 10-12 keV source
falls from around $\sim$10$''$ to $\sim5''$ and after 00:13:00 rises from $\sim$6$''$ to $\sim$13$''$. The 15-20 keV
source also shows this pattern except there is a larger peak at 00:05:00 and then a more pronounced
decrease in width until 00:12:00-00:13:00 UT.

{\it Flare 3-} Again, the change in loop width with time follows a similar pattern as Flares 1 and 2. For all three energies
ranges considered, the loop width decreases from 23:42:00 to the peak in X-ray emission at 23:50:00.
The loop width then increases after this time. The 14-16 keV and 16-20 keV sources $\sim$ fall
from 7-8$''$ at 23:42:00 to $\sim$5$''$ at 23:50:00 and then increase up to 8-9$''$ at 23:58:00.
The 20-25 keV source falls from $11''$ to $5''$. Note the missing data at the fourth and sixth time bins for the 20-25 keV energy range, as VIS FWDFIT was unable to fit successfully at these times for this energy range.

\subsection{Loop length}

As with loop width, each flare shows general pattern for emitting loop length changes with time. Before a peak in X-ray emission,
the source length for each flare tends to decrease and after a peak in X-ray emission, the source length either
increases slightly or remains approximately constant within the errors.

{\it Flare 1 - } For all energies plotted, there is a rapid decrease in source length until the peak X-ray emission
at $\sim$14:30:00-14:32:00. After the peak, the source length remains approximately constant (within the error).
The smallest decrease in loop length before the X-ray peak emission occurs for the 10-12 keV source which falls
from $54''$ to $38''$. The decrease in loop length grows with energy and the highest drop in loop length occurs
for the highest energies plotted at 15-20 keV, which fall from $74''$ to $38''$ before the peak in X-ray emission.

{\it Flare 2 - } The pattern for length changes with time are similar to that of the width changes. For the 10-12 keV source,
the length before the 00:03:00 peak in the lightcurve, falls from $\sim$26$''$ to $\sim$20$''$, rises
to $\sim$24$''$ at 00:11:00, falls to $\sim$20$''$ at 00:15:00 and then increases to $21''$ at 00:17:00. Due to the multiple peaks, the loop length tends to increase after a peak in the lightcurve.

{\it Flare 3 - } For all energies, the length drops rapidly between 23:42:00 and 23:50:00 from the range of $25''$ to 30$''$
to $\sim$10$''$ for all energies plotted. Again as with Flare 1, after the X-ray peak the length of the loop remains
approximately steady until the final plotted time of 23:58:00 for all energies.

\subsection{Loop radial position}

For all three flares, peaks in each X-ray lightcurve tend to denote times where the trends in loop radial position change.

{\it Flare 1- }Since this is a limb event,
the radial position can tell us whether the source is moving away or towards the limb. Before the X-ray emission peak
at 14:30:00-14:32:00, the source moves towards the limb, falling a distance of $\sim2''$ at 10-12 keV, 12-15 keV and 15-20 keV.
Then after the peak, the source moves away from the limb. By plotting the actual source positions,
we also see that overall the entire loop structure moves northwards during the time interval of 14:22:00-14:40:00, moving in a U-shape as also shown for a number of flares in \cite{2008ApJ...686L..37S}. From our 
results for Flare 1, we see that the changes in radial position with time are $\sim$comparable to the width changes and smaller than the length changes. The largest change in position is only $2-3''$ while the width decreases by $3''$ and increases by $6''$.
The length shows the largest change with a decrease of $\sim17''$ or more before the peak in X-ray emission.

{\it Flare 2 - }Overall, the radial positions of the 10-12 keV, 12-15 keV and 15-20 keV sources increase with time.
At points of peak X-ray emission, that is, at $\sim$00:03:00 and $\sim$00:12:00, we see changes in the gradient.
The slope steepens between $\sim$00:03:00 and 00:12:00. Since this is a disk event, it is difficult to say whether there is a change
in source altitude at these peaks (as for Flare 1). By plotting the source positions on the solar disk with time, we can see that
before $\sim$00:03:00 the source moves in a north-easterly direction, it then moves southwestward
until $\sim$00:12:00, before moving northeastwards after this peak. At 10-12 keV, between 23:58:00
and 00:20:00 the source radial direction changes by 9.5$''$. Therefore the overall change in position is larger
 than the individual changes in loop width and loop length for Flare 2.

{\it Flare 3 - }For all energies, the radial distance falls with increasing time. There does not seem to be any significant
difference in the radial distance trend after the X-ray emission peak at 23:50:00, apart from the steadier decrease in radial distance
after this time. As with Flare 2, Flare 3 is a disk source and hence it is difficult to determine if there is any altitude change.
Plotting the $(x,y)$ disk position shows us that the source centroid moves in a north-easterly direction until the peak
at $\sim$23:50:00 and then begins to move in a south-easterly direction, once there is a decrease in the X-ray emission.
For all three energies, the radial distance decreases by $\sim$8$''$ between 23:40:00 and 23:58:00.

\begin{figure*}
\vspace{-15 mm}

\includegraphics[scale=0.35]{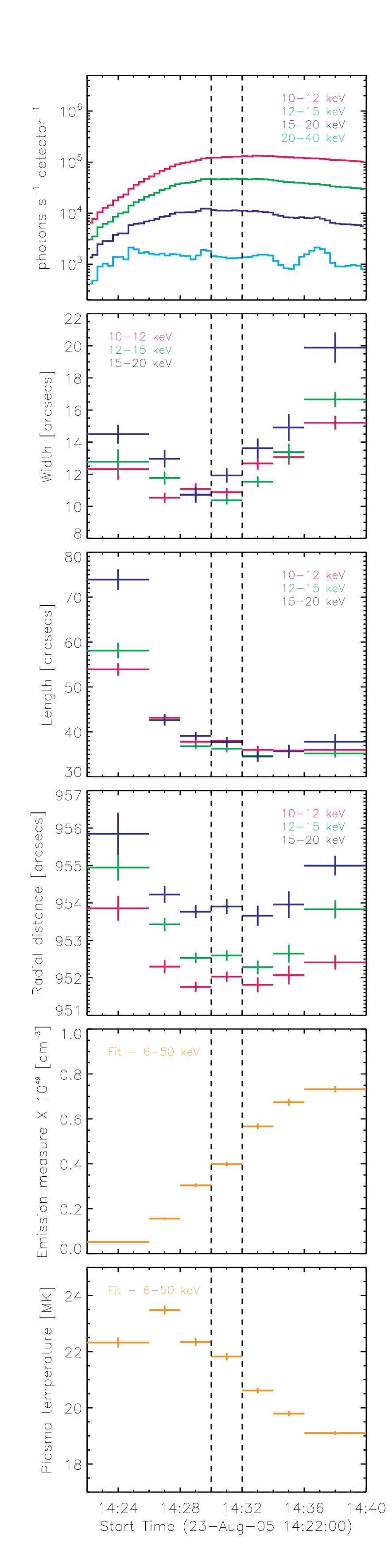}
\includegraphics[scale=0.35]{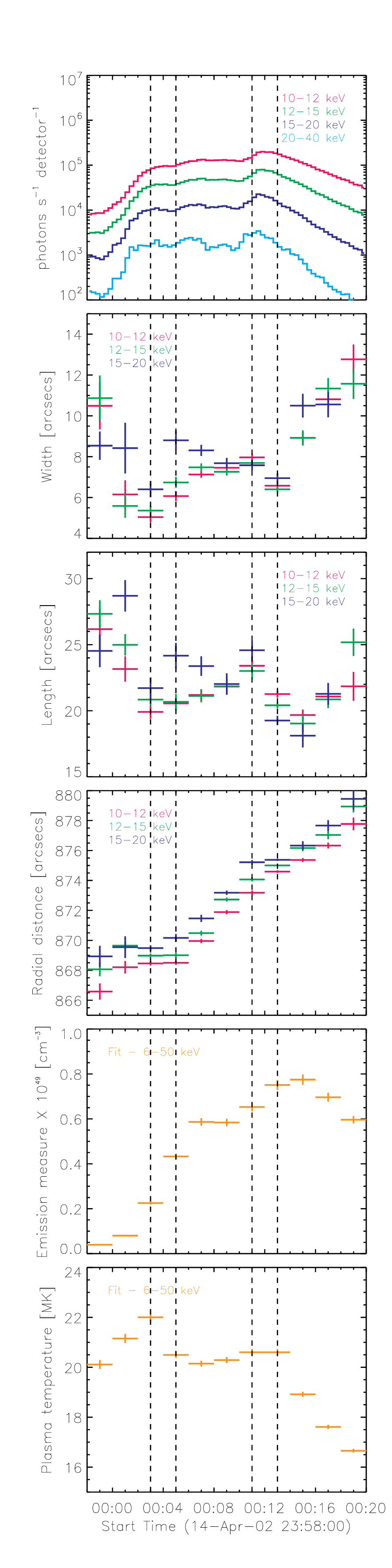}
\includegraphics[scale=0.35]{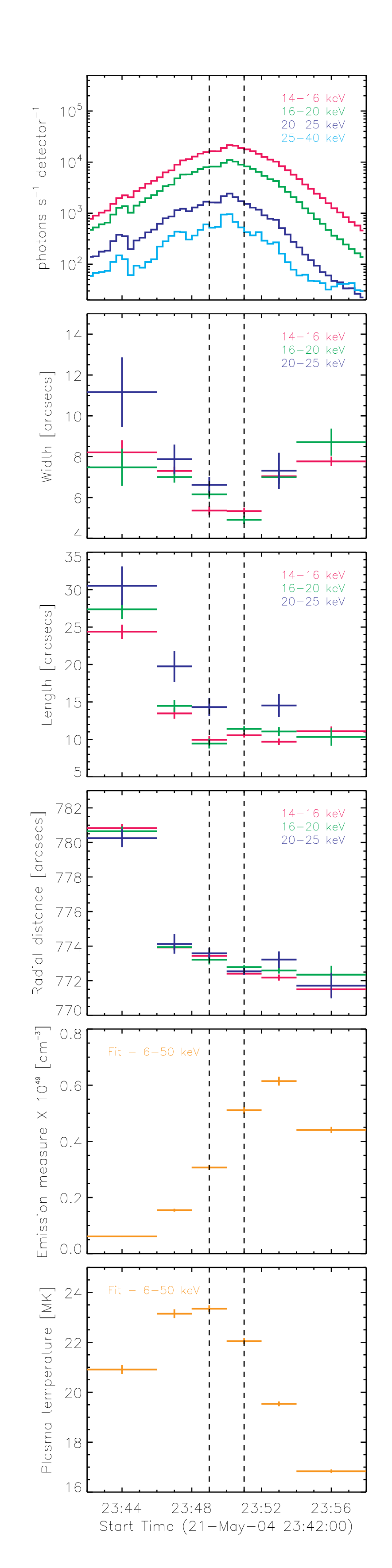}
\caption{(left) 23-August-2005,
(middle) 14-April-2002 and (right) 21-May-2004. 1st row - lightcurves (see graph legend for energies), 2nd row - loop width vs. time,
3rd row - loop length vs. time, 4th row - radial position vs. time, 5th row - emission measure vs. time and 6th row - plasma temperature vs. time.
Dashed lines = peak X-ray emission.}
\label{fig:is_paras}
\end{figure*}

\section{Changes in emitting volume and other inferred parameters}

\begin{figure*}
\vspace{-15 mm}
\includegraphics[scale=0.38]{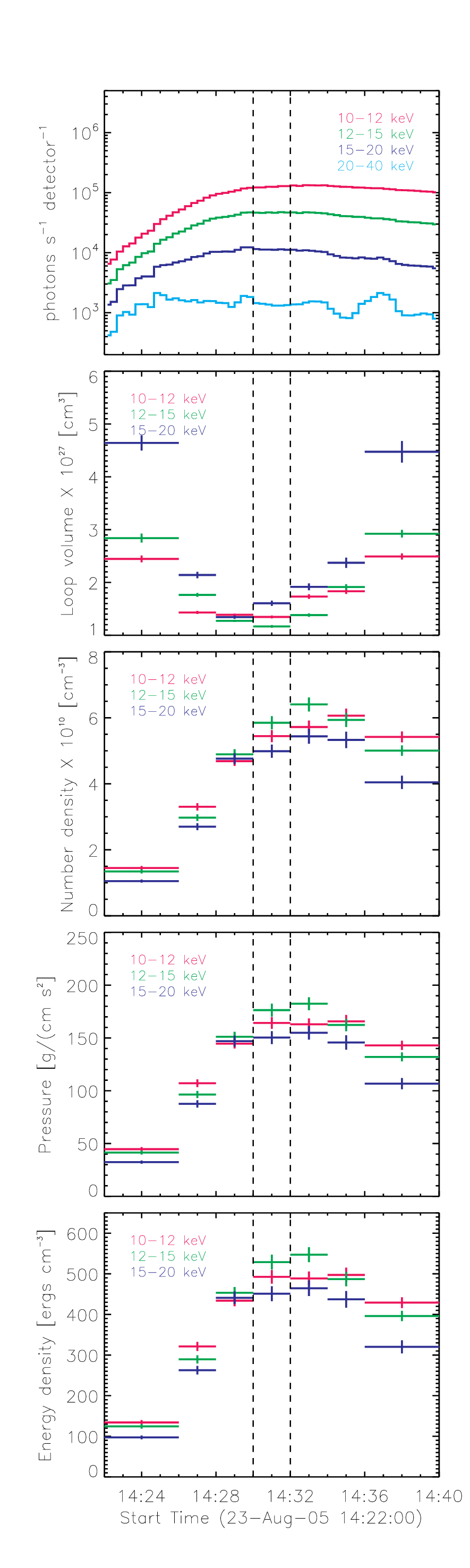}
\includegraphics[scale=0.38]{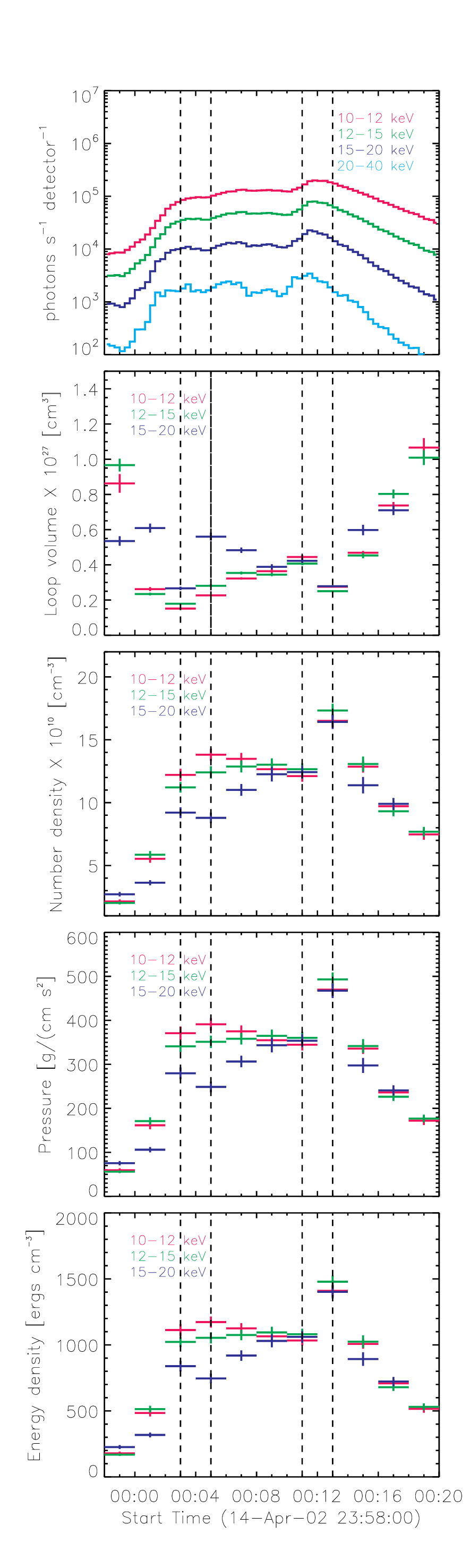}
\includegraphics[scale=0.38]{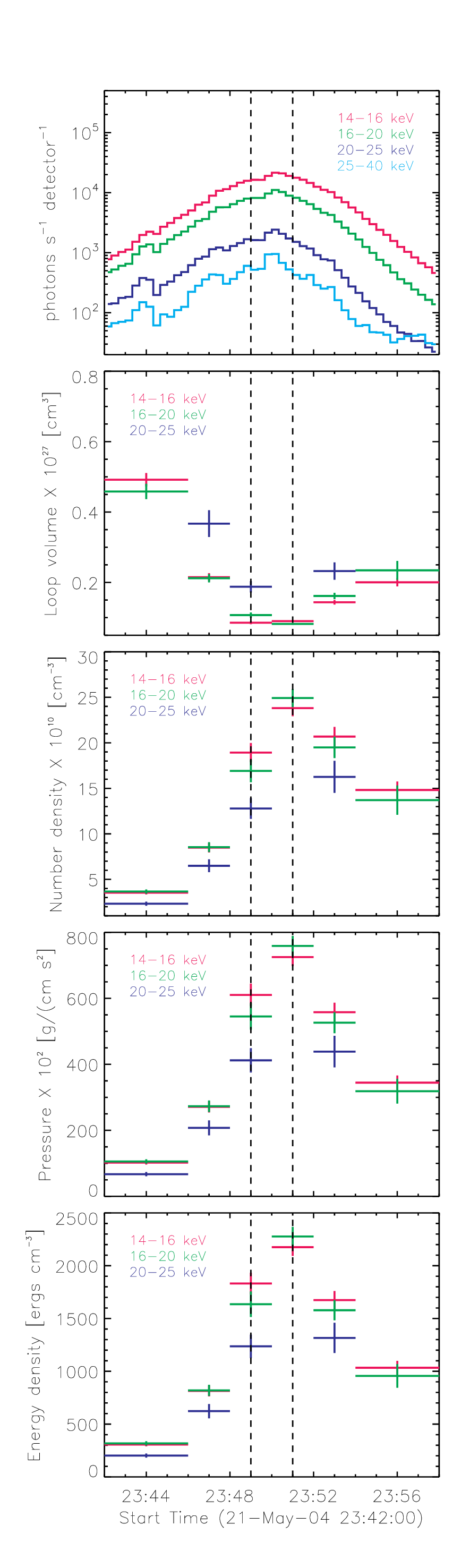}
\caption{(left column) 23rd August 2005,
(middle column) 14th April 2002 and (right column) 21st May 2004. 1st row - Lightcurves for each event, 2nd row - Volume changes with time,
3rd row - Number density changes with time, 4th row - Thermal pressure changes with time and 5th row - thermal energy density changes with time.}
\label{fig:com_paras}
\end{figure*}

From the loop width FWHM $W$ and loop length FWHM $L$, we can infer the general changes in the emitting loop volume, $V$,
for each flare over time, at each energy band. We assume that the volume of the loop is given by 
$V=\pi W^{2}L/4$ that is, assuming a cylindrical loop. The changes in emitting loop volume with time for Flares 1, 2 and 3 are plotted
in Figure \ref{fig:com_paras} (second row). In general, for all three events, the source volume decreases before a peak
in X-ray emission and increases after a peak in X-ray emission. The changes in plasma number density, thermal pressure and thermal energy density can all be calculated using combinations of loop volume, emission measure
and plasma temperature. The plasma number density, $n$, can be obtained via
$n=\sqrt{EM/V}$, the pressure, $P$, from $P=nk_{B}T$, where $k_{B}$ is the Boltzmann constant
and finally the energy density, $E=3nk_{B}T$. The variation of these quantities with time for Flares 1, 2 and 3
are shown in Figure \ref{fig:com_paras}  (third, fourth and fifth rows respectively).

{\it Flare 1 - }As expected from the width and length results,
the loop volume falls between 14:22:00 and the peak in the X-ray lightcurve at $\sim$14:30:00 and then rises
after this time for all three energies. For all three energy bands, the decrease and increase in loop volume occurs at roughly the same rate. The largest decrease is for the highest 15-20 keV band, falling from $\sim4.7\times10^{27}$ cm$^{3}$ to $\sim1.4\times10^{27}$ cm$^{3}$ at 14:29:00 and then rising after this time to $\sim4.7\times10^{27}$ cm$^{3}$ at the last observational time. The 10-12 keV band falls from $\sim2.4\times10^{27}$ cm$^{3}$ to $\sim1.3\times10^{27}$ cm$^{3}$ at 14:31:00 and then rises back to $\sim2.5\times10^{27}$ cm$^{3}$ at the final observational time. The number density, thermal pressure and energy density for all three energy bands tend to follow 
the same pattern, rising to a peak at some time after the peak X-ray emission and then slowly decreasing. For the 10-12 keV band, the number density rises from $1.5\times10^{10}$ cm$^{-3}$ at 14:22:00
to $6.5\times10^{10}$ cm$^{-3}$ at 14:35:00. It then falls to $\sim5.6\times10^{10}$ cm$^{-3}$ at 14:38:00. The 12-15 keV and 15-20 keV bands follow similar patterns, peaking at $\sim$14:33:00.
The pressure rises from $\sim40$ g/[cm s$^{2}$] at 14:22:00
and reaches $170$ g/[cm s$^{2}$]  at 14:31:00, the time where the X-ray emission peaks. After this time the pressure
remains approximately constant at $\sim160-170$ g/[cm s$^{2}$] until the last observation time where it falls to $\sim140$ g/[cm s$^{2}$]. The 12-15 keV and 15-20 keV energy bands peak at $\sim$14:33:00.
The energy density is just the pressure multiplied by 3 and hence it follows the same pattern as pressure throughout the observed duration of the flare.
The energy density peaks at $500$ ergs cm$^{-3}$ in the 12-15 keV band at $\sim$14:33:00, after the peak X-ray emission.

{\it Flare 2 - }For the 10-12 keV and 12-15 keV sources, the loop volume falls until the first peak at 00:03:00
and then increases until it reaches 00:12:00, drops at 00:13:00 and then increases again after this time.
As for Flare 1, the number density, thermal pressure and energy 
density all follow the same pattern for each energy band, rising to a time at or just after the peak in X-ray emission and then decreasing after this point. The highest number density, thermal pressure and energy density for each energy band occurs at 00:12:00 to 00:14:00, just after the peak in X-ray at 00:11:00-00:13:00. The number density peaks at around $18\times10^{10}$ cm$^{-3}$ in the 12-15 keV band, while the thermal pressure and energy density peak at $\sim500$ g/[cm s$^{2}$] and $1500$ ergs cm$^{-3}$ respectively at this time and energy.

{\it Flare 3 - }The loop volume again falls before the peak in X-ray emission at 23:50:00
and then rises again after this time, for all three energies.
For this flare again, the number density, thermal pressure and energy density all peak just after the peak in X-ray emission. At 23:50:00-23:52:00 the number density peaks at $23\times10^{10}$ cm$^{-3}$, the thermal pressure peaks at $750$ g/[cm s$^{2}$] and the energy density at $\sim2300$ ergs cm$^{-3}$, in the 16-20 keV band.

\section{Summary and discussion}

Using visibility forward fitting, we have for the first time produced a dedicated study of how the spatial and spectral properties
of three coronal X-ray loops changed during the temporal evolution of each flare. All X-ray loops exhibited similar changes in both their spatial and spectral properties and hence the results indicate that a common process is
occurring for all three events; the emitting flaring loop volume is decreasing before the peak in X-rays and increases after the peak in X-rays is reached.
Before the peak X-ray emission, the emitting lengths and widths of each coronal loop decreased with time, indicating that the X-ray emitting region of the loop volume 
was contracting (a reduction in loop corpulence and length) as the X-ray emission from the region grew. After the X-ray peak, the loop corpulence increased at approximately the same rate as during contraction stage. For Flares 1 and 3, with one peak in their lightcurves, once the minimum X-ray loop length was reached during the X-ray peak it remained approximately constant, at least within the errors of our results.

Similar to previous studies \citep{1978ApJ...220.1137A,1980ApJ...242.1243S,1984ApJ...285..835G,1999ApJ...514..472M}, 
spectroscopy for each event showed that the plasma temperature initially grew but began to decrease before the peaks in X-ray emission and emission measure. The emission measure for each flare generally grew with dips observed during the final observational times of Flares 2 and 3. At the same time, the number density, thermal pressure and energy density of the plasma 
also increased as the X-ray emission grew.  The plasma temperature decreased much slower than by thermal conduction only, even during the X-ray decay phase, suggesting at later stages additional energy 
release is required \citep[e.g.][]{2011A&A...531A..57K} to explain the long lasting X-ray loop emission.
For the limb event (Flare 1), we do see a decrease in loop altitude before the peak in X-ray emission of roughly $2''\simeq1.4$ Mm, which is comparable in magnitude to the decrease in loop width but overall, the largest changes occurred for the X-ray loop volume, not the X-ray loop position. Decreases in loop altitude before the peak in X-ray emission have been well noted before and are often referred to as coronal implosion, loop contraction or field line shrinking.

\cite{2004ApJ...612..546S} also studied the loop position changes of Flare 2. From their observations, they did conclude 
that the loop was indeed decreasing in altitude before the first peak in the lightcurve at 00:04:00.
They concluded that the loop decreased by $\sim2''$ over a 4 minute period, which is consistent with our radial distance results for Flare 2. However, no altitude decrease before the larger peak in X-ray emission at 00:12:00 was noted, where again we see decreases in loop length and loop width.

\begin{figure}
\includegraphics[scale=0.46]{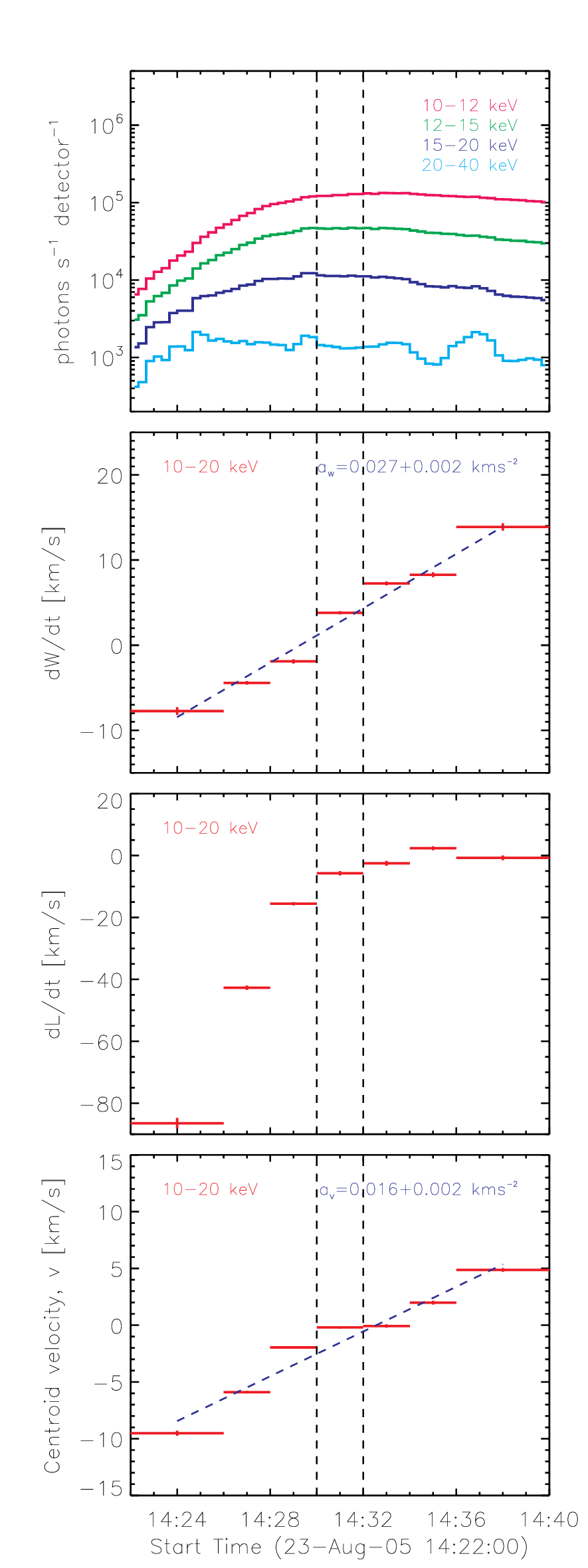}
\caption{For Flare 1 (23rd August 2005), lightcurve (top row), $dW/dt$ (2nd row), $dL/dt$ (third row) and $dr/dt=v$ (bottom row). Both $dW/dt$ and $v$ can be
fitted using a straight line indicating constant accelerations $a_{W}$ and $a_{v}$ and hence constant forces.}
\label{fig:dvwl}
\end{figure}

\begin{figure*}
\hspace{-13 mm}
\includegraphics[scale=.40]{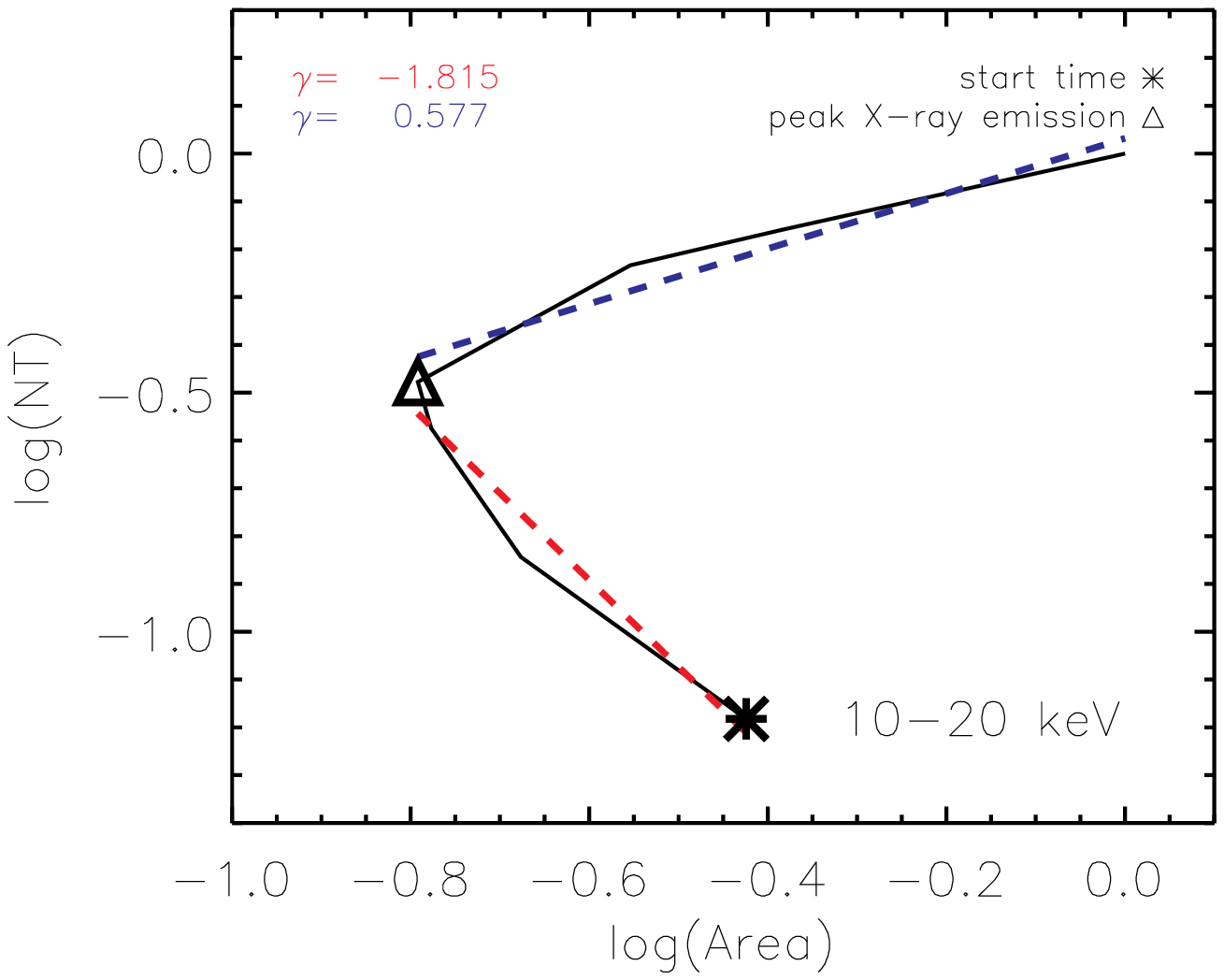}
\hspace{-5.4 mm}
\includegraphics[scale=.40]{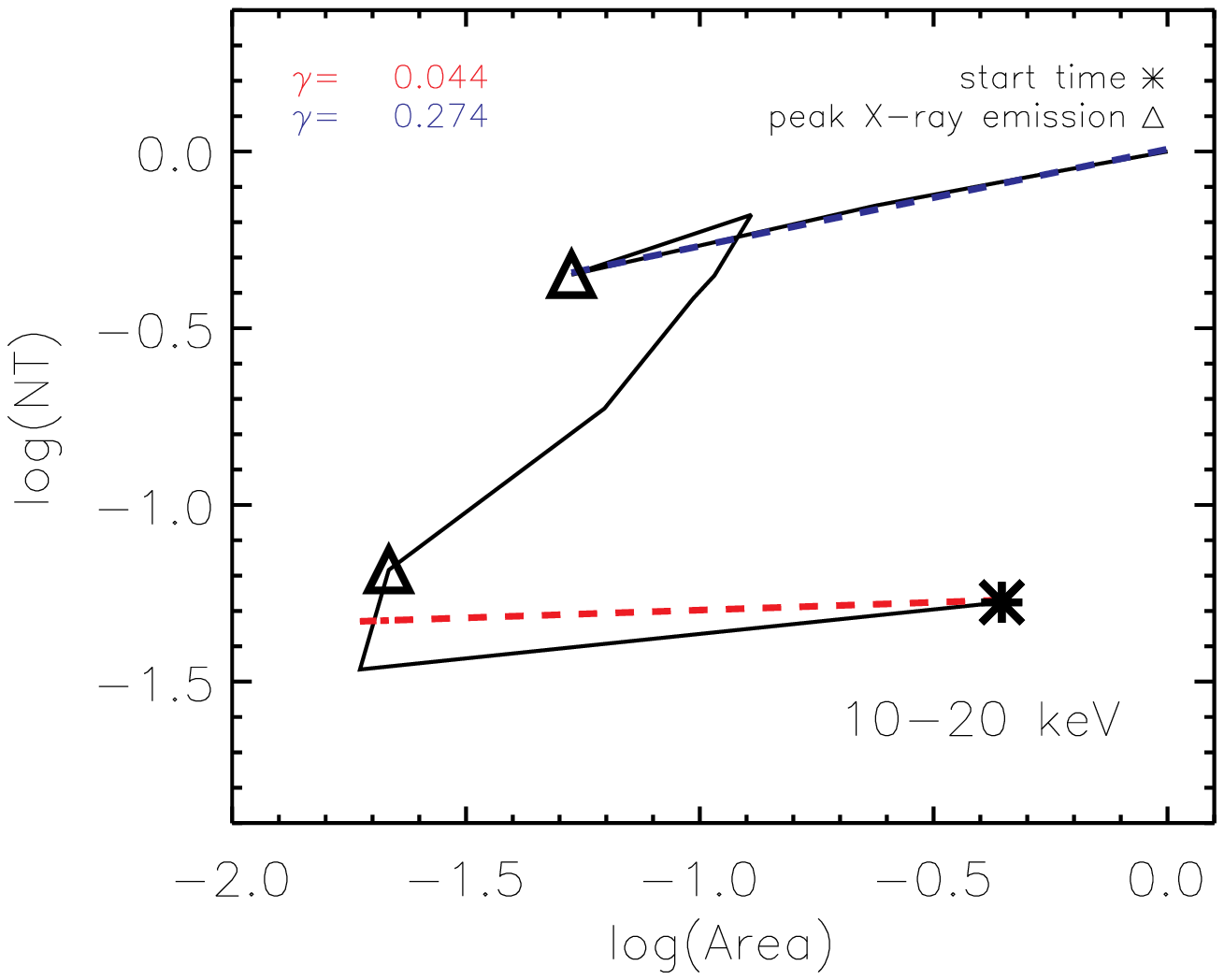}
\hspace{-5.4 mm}
\includegraphics[scale=.40]{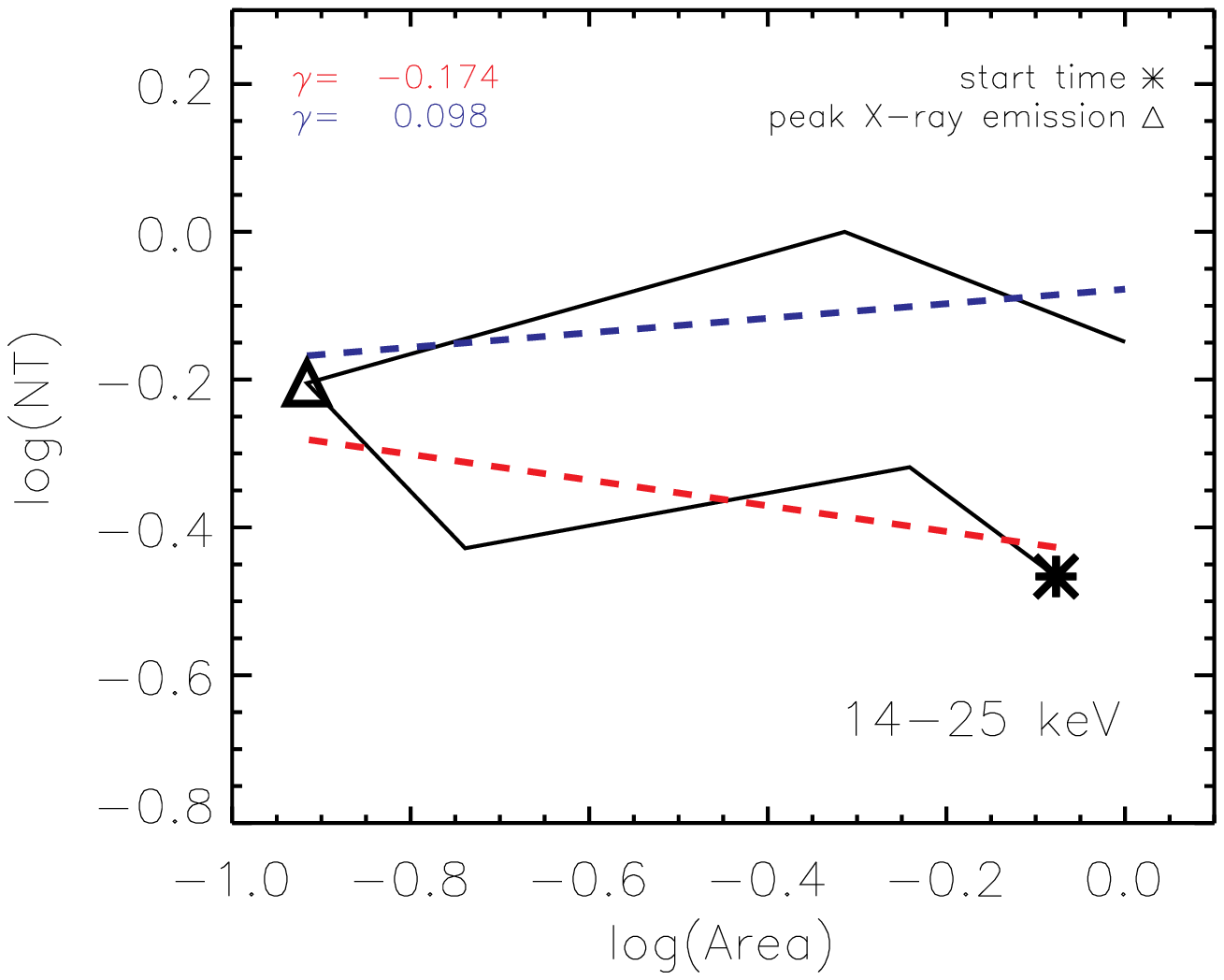}
\hspace{-13 mm}
\caption{Plots of $\log{NT}$ against $\log{1/A}$ for Flares 1 (left), 2 (middle) and 3 (right).  The star represents the start time
of each event while the triangle represents the peaks in X-ray emission for each event. The red and blue
dashed lines represent straight line fits during the compressive (red) and expansive (blue) phases.
For Flare 2, only the first compressive and final expansion phase have been fitted. For simple compressive
heating, we would expect the gradient of each line to be $\gamma=-1$.}
\label{fig:NTA}
\end{figure*}

It has been suggested that a decrease in loop altitude may be an indication of collapsing magnetic trap acceleration/heating \citep{1966PASJ...18...57T,1997ApJ...485..859S,2002A&ARv..10..313P,2004A&A...419.1159K,2006A&A...446..675V,2012A&A...546A..85G}.
Although this may be the case for other events, we do not believe that the collapsing magnetic trap is the prime solution
to our observations, firstly due to overall larger loop volume changes over relatively small position changes.
\cite{2006A&A...446..675V} observed the coronal loop of a GOES X-class flare and found downward velocities as large 
as $\sim14$ km/s in 10-15 keV band and $\sim29$ km/s in the 15-20 keV band prior to the peak in X-ray emission. 
This is not what we observe for these events. For Flare 1, we only observe an average downward velocity in the 10-20 keV 
band of $\sim4$ km/s, which is generally comparable to the decrease in loop corpulence during this period. 
Average changes in loop radial position for Flares 2 and 3 are also $\sim4$ km/s. More convincingly, in a collapsing magnetic 
trap model, simple compressive heating during the contraction stage would imply that
$NT\propto1/A$ \citep{1978ApJ...223.1058M,1981ApJ...244..653E}, where $N$ is the number of particles in the region, 
$T$ is the plasma temperature and $A$ is the cross-sectional area of the region, given by $\pi W^{2}$, where $W$ is the loop width. 
Figure \ref{fig:NTA} plots $\log{NT}$ against $\log{A}$ for each flare and shows this not to be the case.
Straight lines fits to both the contraction and expansion phases show 
that the gradients are either greater or less than $-1$. 

\begin{figure*}
\epsscale{2.0}
\plotone{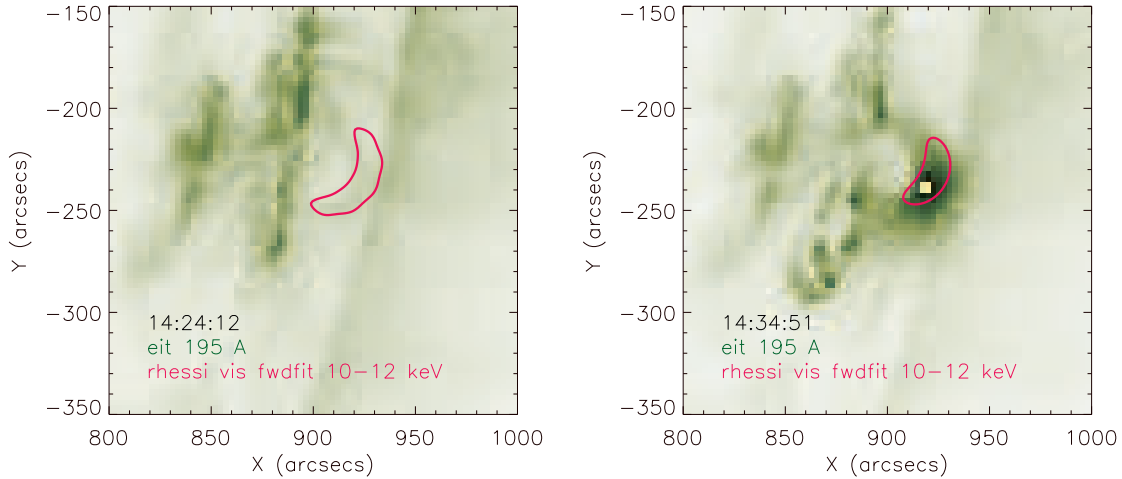}
\caption{SOHO EIT $195\AA$ images for Flare 1 at the times of 14:21:12 and 14:34:51, corresponding 
to the times of rise and peak in X-ray emission. RHESSI 10-12 keV X-ray contours are over plotted in pink. 
Note the lack of bright EUV emission at the rise stage of this flare.}
\label{fig:fig_eit}
\end{figure*}

Observations by \cite{2011ApJ...730L..22K} showed how the loop corpulence of the 14/15th April 2002 event (Flare 2) increased with
energy during two times corresponding to the first rise and first decay stages of our observations and suggested that the presence of magnetic 
turbulence (the diffusion of field lines) was the cause of the energetic width increases. Hence, the suggestion that magnetic
turbulence in the region can account for the energetic changes in loop corpulence, may also be able to account for the extra
energy in the loop, since both \cite{2011ApJ...730L..22K} and \cite{2011A&A...535A..18B} found the energy density of 
magnetic fluctuations to be significant and could be comparable to that of the flaring plasma and higher than the energy density of non-thermal particles. Flares 1 and 3 also show similar length and width increases with energy at a single time range, also suggesting the presence of magnetic turbulence. 

Many observations \citep{2005ApJ...629L.137L,2009ApJ...696..121L,2009ApJ...706.1438J,2010ApJ...724..171R,2012ApJ...749...85G} of loop height and length changes have been explained in terms of a reduction in magnetic
pressure. Usually the reduction in magnetic pressure is referred to as Taylor relaxation but this only refers to a special case where the resulting field is linear force free \citep{1974PhRvL..33.1139T}. The reduction in magnetic pressure could also account for the reduction of loop corpulence 
or cross-sectional area, as shown in simulations by \cite{2007A&A...472..957J} and hence the observed trends of number density and pressure.
\cite{2009ApJ...696..121L} studied coronal implosion of one coronal source. They explained the reduction in height of a coronal source
before the peak in X-ray emission in terms of Taylor relaxation. They also suggested that this type of event can only occur
if the coronal loop is already filled with hot, dense plasma before the onset of new event, that is from a previous event in that region.
The fact we do not see EUV emission during the rise phase of Flare 1 seems to correlate with the observations
and suggestions of \cite{2009ApJ...696..121L}. Figure \ref{fig:fig_eit} shows SOHO (Solar and Heliospheric Observatory)
EIT (Extreme Ultraviolet Imaging Telescope) 195$\AA$ images at the times of 14:21:12 and 14:34:51 for Flare 1.
X-ray emission contours at 10-12 keV for 14:22:00-14:26:00 and 14:34:00-14:36:00 are also over plotted.
From Figure \ref{fig:fig_eit}, we can see that during the rise phase there is no bright 195$\AA$ EUV emission emanating from the loop,
only 10-20 keV X-ray emission. After the peak in X-ray emission, we do see EUV emission from the loop.
We see an overall increase in the number of particles, $N$, within the loop region throughout the 
duration of all three events. Due to low coronal densities, chromospheric evaporation probably accounts for this increasing $N$, initially driven by thermal conduction and possibly at later times 
by electrons reaching the chromosphere, when we see weak footpoint emission and EUV emission from the loop.

\subsection{Three temporal phases and suggested explanations for the observations}

\begin{figure*}
\vspace{-20 mm}
\includegraphics[scale=.38]{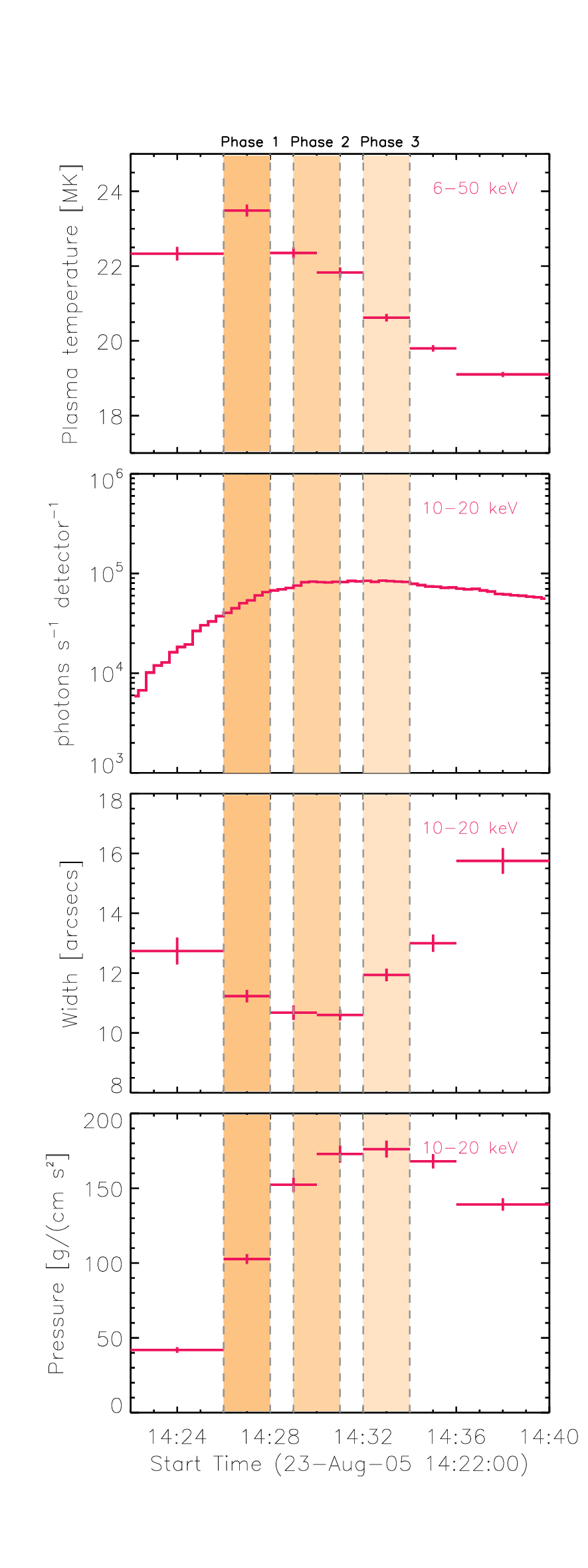}
\includegraphics[scale=.38]{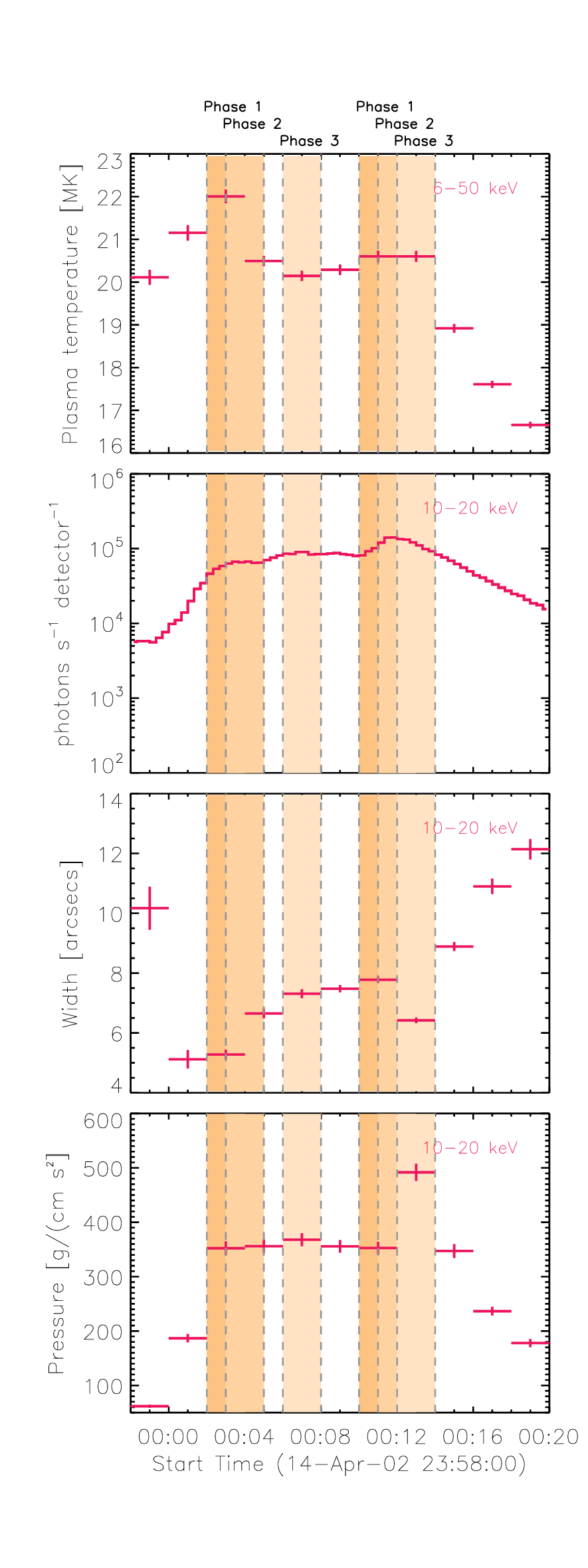}
\includegraphics[scale=.38]{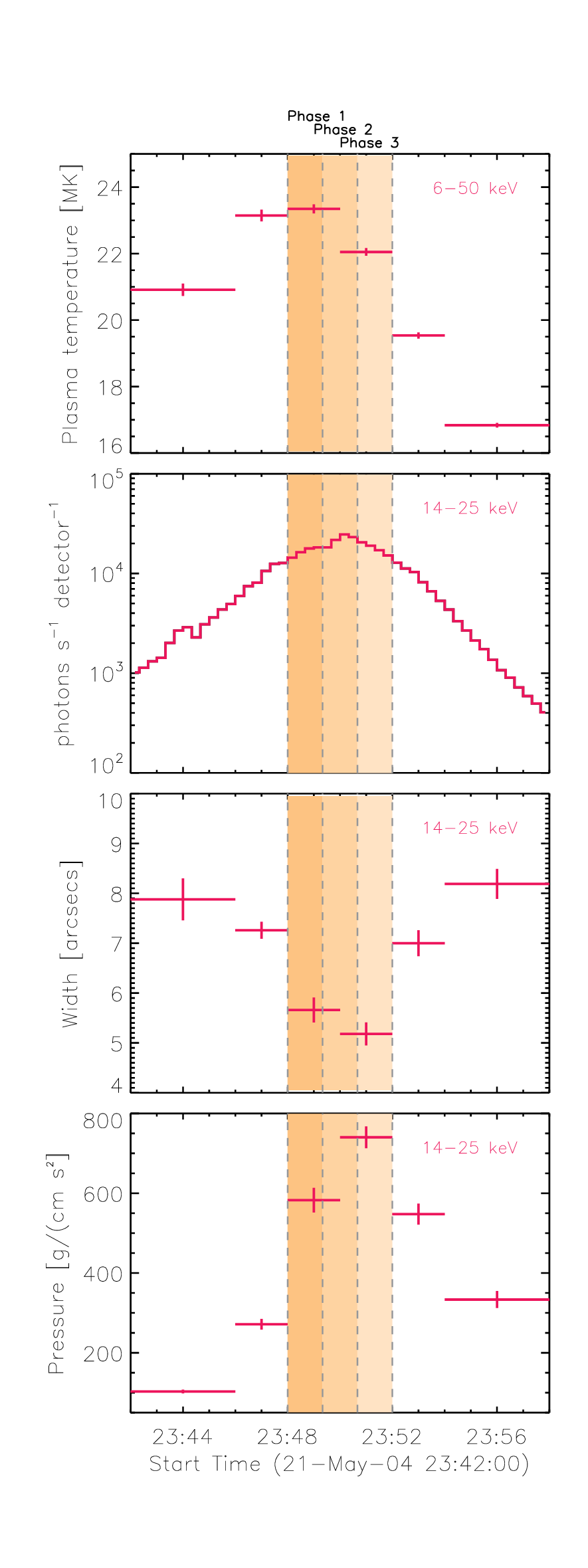}
\caption{Observations of plasma temperature, X-ray emission, loop width and thermal pressure are replotted together for Flares 1 (left), 
2 (middle) and 3 (right) at one energy band of 10-20 keV (14-25 keV for Flare 3). The orange bars represent three phases: 1. a peak in plasma temperature, 2. a peak in X-ray emission, generally coinciding with the smallest loop width and 3. a peak in thermal pressure.}
\label{fig:modphases}
\end{figure*}

For each flare, Figure \ref{fig:modphases} replots the plasma temperatures, X-ray emissions, emitting loop widths and thermal pressures but now at a single energy band of 10-20 keV (14-25 keV for Flare 3). From Figures \ref{fig:is_paras} and \ref{fig:com_paras} and now more clearly in Figure \ref{fig:modphases}, we can see that collectively, the observations of each flare display three distinct phases and each of these phases will form the basis of our suggested explanation. Each phase is represented by a shaded orange bar. During Phase 1, there is a peak in plasma temperature and during Phase 2, a peak in X-ray emission. At Phase 2 we see the smallest emitting loop width, length and hence volume. Finally, during Phase 3, the thermal pressure of the region peaks. For Flare 1 (left column), each phase is well separated and can be clearly seen. The pattern can also be seen for Flares 2 (middle column) and 3 (right column) but each phase is not as clearly defined as for Flare 1. During Flare 3, each phase occurs over much shorter time intervals and therefore each of the phases overlap slightly. During Flare 2, there are multiple peaks, which along with the shorter timescales for each process, makes each individual phase harder to see. However, the overall pattern is observed for all three flares.
We can see from Figure \ref{fig:modphases} that each phase can only be easily seen for slower events. The quicker the event, the harder it is to distinguish between each of the three phases as each phase overlaps in time. Many solar eruptive events may exhibit a similar pattern but each phase may occur at timescales too fast to be currently observed.

In order to understand our observations, we believe it is crucial to understand the decreasing X-ray widths before the X-ray peak. Length variations along the field can always be explained in terms of changing number density, but in order to explain the corpulence variations, where electrons are tied to the magnetic field, the most plausible explanation is the movement or the diffusion of magnetic field lines. We have already noted the energetic loop width increases at a given time for each flare, suggesting the presence of turbulence in the region, but overall in time the width at a given energy band shrinks until the X-ray peak is reached, implying the presence of both turbulence and the shrinking of the cross-sectional area of the field at this stage of the flare. Therefore, we suggest the plasma within our emitting X-ray loop region is tied to the magnetic field lines and hence the contraction of the emitting loop cross-section or corpulence during the rise phase is ultimately due to the contraction of the cross-sectional area of field lines that thread the region or possibly the expansion of field lines above the region. It is also sensible to assume that the region we call the loop in X-rays actually consists of multiple coronal loops that cannot be resolved in X-rays using RHESSI. Though we can only see what is happening to the region of the loop that is emitting X-rays, we will assume the entire loop region is contracting since the plasma is tied to the field lines. Hopefully, from future observations we will find similar, newer coronal events with complimentary AIA/SDO data where it may
be possible to see if the region consists of multiple coronal loops. From our observations, the reason for this cross-sectional width shrinking is unknown and beyond the scope of this paper, but we can speculate for further study that it may be due to a reduction in magnetic pressure within the X-ray loop region, as often suggested to describe height decreases (coronal implosion or loop contraction) discussed earlier \citep{2005ApJ...629L.137L,2009ApJ...696..121L,2009ApJ...706.1438J,2010ApJ...724..171R,2012ApJ...749...85G}. The magnetic pressure decreases as the magnetic field relaxes from a non-potential state. Although it is reasonable to assume that during a solar eruptive event such as a flare, the field will reside in a non-potential state, X-ray observations of these events do provide evidence for the non-potentiality, again through the inference of magnetic turbulence and hence a non-parallel field component within the loop region. The non-potential state and decreasing magnetic pressure could be due to reconnection above the region or maybe even along the loop itself \citep{2004ApJ...608..540V,2011ApJ...729..101G,2012SoPh..277..299G,2012SoPh..tmp..225G}. One such model is caused by a kink instability as shown in recent simulations by \cite{2011ApJ...729..101G,2012SoPh..277..299G,2012SoPh..tmp..225G}. In this model, the energy of twisted loops is released by reconnection inside the loop and transferred to plasma heating and particle accelerations.

During phase 1, the process causing the contraction of loop width/cross-section is also probably responsible for the comparable decrease in loop altitude, or at least the change in loop position observed for Flares 2 and 3 sitting on the solar disk. Plots of $dW/dt$ (width contraction/expansion) and $dr/dt=v$ (centroid velocity) for Flare 1 plotted in Figure \ref{fig:dvwl} show that each parameter can be fitted using a straight line during both the contraction and expansion phases. $dL/dt$ (length contraction/expansion) is also plotted in Figure \ref{fig:dvwl} but its trend cannot be described by a straight line fit suggesting the changes in length are caused by a different process, itself a consequence of the process producing the decreasing loop corpulence. It is also sensible to assume that this process is responsible for the increased plasma temperature and the acceleration of electrons. 
The increasing temperature of the region means that energy will be thermally conducted towards the lower levels of the solar atmosphere causing gentle chromospheric evaporation of the denser coronal and chromospheric layers below. The observations show the plasma temperature peaking relatively early before the peak in X-ray emission and then slowly decreasing; slower than by thermal conduction, implying that energy is still being supplied to the loop plasma. This is most likely via the conversion of magnetic energy. Chromospheric evaporation drives plasma into the region producing the increasing number density and hence thermal pressure, along with the shrinking corpulence at this phase. The increasing number density is responsible for the rapid non-linear decreasing X-ray loop length, since electrons accelerated within the region will travel shorter distances before interacting.

{\it Phase 2} During Phase 2, after the peak in plasma temperature, the loop corpulence stops shrinking. Number density and thermal pressure within the loop still increases due to chromospheric evaporation and the length of the emitting region also reaches its lowest point. The loop corpulence may stop shrinking because the process causing the shrinking ceases or it may be due to the balancing of forces within the region. For example, if the reduction of loop corpulence is due to the reduction of magnetic pressure in the loop then at this phase, the growing thermal pressure may finally be high enough to balance the reduction in $B$ pressure.

{\it Phase 3} During the final phase, the thermal pressure continues to rise within the loop due to the increasing number density from chromospheric evaporation. We believe the growing thermal pressure in the region is now responsible for the expanding loop corpulence. This expansion, in turn, eventually halts the increasing number density and thermal pressure at a time after the peak in X-ray emission. After Phase 3, the loop corpulence continues to increase and we see slow decreases in both number density and thermal pressure. The X-ray emission continues to decrease during and after Phase 3 and the emitting loop length during this period remains approximately constant, equal to the minimum loop length in Phase 2, even with a decreasing number density. It is sensible to assume that the acceleration mechanism in the loop is slowing during this time. However, Flare 2 is an exception to this trend as we see multiple events/X-ray peaks in the lightcurve.

Finally, we would like to note that a study of the 23rd July 2002 flare by \cite{2010ApJ...725L.161C} calculated temporal volume changes using the Clean algorithm and assuming an elliptical geometry. Overall this flare shows a general trend consistent with our results; an overall decrease in volume before the peak in X-ray emission and an overall increase in volume after the peak in X-ray emission. This flare also shows a peak in plasma temperature before the peak in X-ray emission and a high number density (and hence thermal pressure) after the X-ray emission first peaks.

 \acknowledgments

NLSJ is funded by STFC and SUPA scholarships. EPK gratefully acknowledges the financial support by STFC
Rolling Grant, and by the European Commission through the HESPE (FP7-SPACE-2010-263086) Network
acknowledged.


\end{document}